\documentclass[twocolumn,showpacs,preprintnumbers,amsmath,amssymb]{revtex4}

\usepackage{graphicx}
\usepackage{dcolumn}
\usepackage{bm}

\begin{document}

\title{Phenomenological theory of magnetization reversal 
in nanosystems with competing anisotropies 
}

\author{A. A. Leonov$^{1,2}$,  U.K. R\"o\ss ler$^{1}$, A.N.\ Bogdanov$^{1}$}
\affiliation{$^1$IFW Dresden, Postfach 270116, D-01171 Dresden, Germany}
\affiliation{$^2$Donetsk Institute for Physics and Technology, 
R. Luxemburg 72, 83114 Donetsk, Ukraine}


%
\date{\today}

\begin{abstract}
{
The interplay between intrinsic and 
surface/interface-induced magnetic anisotropies  
strongly influences magnetization processes 
in nanomagnetic systems.
We develop a micromagnetic theory to describe
the field-driven reorientation in nanomagnets
with cubic and uniaxial anisotropies.
Spin configurations in competing phases  
and parameters of accompanying multidomain states 
are calculated as functions of the applied field
and the magnetic anisotropies.
The constructed magnetic phase diagrams allow to classify
different types of the magnetization reversal 
and to provide detailed analysis of the switching processes
in magnetic nanostructures.
The calculated magnetization profiles of isolated
domain walls show
that the equilibrium parameters of such walls are extremely
sensitive to applied magnetic field and values of the competing
anisotropies and can vary in a broad range. 
For nanolayers with perpendicular anisotropy 
the geometrical parameters of stripe domains 
have been calculated as functions of a bias field.
The results are applied to analyse 
the magnetization processes as observed 
in various nanosystems with competing
anisotropies, mainly, in diluted magnetic semiconductor films 
(Ga,Mn)As.
}
\end{abstract}


         
\maketitle

%

\vspace{5mm}

\section{Introduction}


In magnetic nanostructures complex physical processes on surfaces
and interfaces give rise to enhanced uniaxial
magnetic anisotropies \cite{Martin03,Johnson96,Bader02,Broeder91}.
The interplay between these induced and intrinsic (magnetocrystalline)
anisotropies strongly influences the magnetization processes 
in many important classes of nanoscaled systems,
such as ferromagnet-antiferromagnet bilayers \cite{bilayers1,bilayers2,bilayers3,Lai01},
thin epilayers of diluted magnetic semiconductors  
\cite{Macdonald05,Sawicki04,Dietl04,
Sawicki004,Sawicki05,Sawicki06}
or in magnetic nanoparticles
\cite{APL07,Duxin00,Jamet04,Wernsdorfer02,Wernsdorfer97}, 
and is the reason of various remarkable effects 
involving complex spin reorientation \cite{Moore03,Liu03,Sawicki04,
Welp04,Wang05,Titova05,Liu05,Sawicki06,311A,311B,Albrecht}
and the evolution of specific multidomain states
\cite{Shono00,Welp03,Fukumura01,Pross04,Thevenard06}.

Most theoretical studies in this field are restricted
to the investigations of \textit{Stoner-Wohlfarth} 
processes through coherent switching \cite{Stoner}
in models with uniaxial and cubic anisotropies
\cite{Torok64,Mitsek1,Mitsek2,Asti80,Millev98,Oepen00,Thiaville98}.
Such theories describe magnetization reversal in a
limiting case of ideally hard magnetic materials.
However, in real magnetic materials 
the reversal processes will usually take 
place by the formation of heterogeneous 
states consisting of the competing phases 
and their transformation under the influence 
of the applied field.
The analysis of such \textit{multidomain} states 
in systems with competing anisotropies 
and their influence on magnetization reversal
is the subject of this paper. 
These investigations can be executed within 
a regular micromagnetic theory \cite{UFN88,Hubert98},
adapted to nanoscaled systems 
(see e. g. Refs.~\onlinecite{PRL01,FTT06,Bogdanov02}).

In section II we introduce the 
phenomenological model and methods;
in the next section we derive 
all possible magnetic configurations in 
the system, calculate their stability
limits, and describe reorientation 
effects \cite{UFN88,JMMM07,JMMM05}.
These results enable us to calculate
the parameters of the multidomain
states and analyse the magnetization 
processes (section IV).
In section V we apply our results
to interpret reorientation effects
and magnetization reversal as observed in
experimental works on nanolayers
of diluted magnetic semiconductors, FM/AFM bilayers,
thin films of Heusler alloys, and magnetic nanoparticles.
In section VI the calculated equilibrium 
parameters of the isolated domain
walls and stripe domains are used
to analyse recent experimental
results in (Ga,Mn)As films with perpendicular
anisotropy.

\section{Phenomenological model}

Within the standard phenomenological theory the magnetic 
energy of a nanoscale ferromagnetic sample 
can be written as a functional 
$W_m=\int\ w({\mathbf r})dV$ with an energy density 
\begin{equation}
w= A\,\sum_{i} \left( \frac{\partial \mathbf m}{\partial x_i} 
\right)^2 - {\mathbf M\cdot \mathbf{H}^{(e)}}\,
-\,\frac{1}{2}\mathbf{M} \cdot \mathbf{H}^{(d)}
+w_a\,,
\label{density}
\end{equation}
where ${\mathbf m}={\mathbf M}/{M_0} \,
(M_0=|\mathbf M|)$ is the normalized value 
of the magnetization vector ${\mathbf M}$, $A$ is 
the exchange constant, and ${\mathbf H^{(e)}}$ and $\mathbf{H}^{(d)}$ 
are the external and demagnetizing fields, respectively.
The anisotropy energy density includes uniaxial anisotropy 
($K_u$) with the axis ${\mathbf a}$ 
and cubic anisotropy ($K_c$) with unity 
vectors $\mathbf n_j$ along cubic axes

\begin{equation}
w_a (\mathbf {M})=-K_u \left(\mathbf {m\cdot a}\right)^2
-\frac{1}{2}K_c {\sum_{j=1}^3} ({\bf m\cdot n_j})^4\,.
\label{anisotropy1}
\end{equation}
The coefficients  $K_u$ and $K_c$ are assumed to
be \textit{positive}. %
Hence, $\mathbf{a}$ and $\mathbf{n}_j$ 
directions are easy uniaxial and
cubic magnetizaton axes, respectively.

The equilibrium 
distribution of the magnetization $\mathbf{M} (\mathbf{r})$
is generally spatially inhomogeneous.
It can be derived directly by solving 
the equations minimizing the energy functional Eq.~(\ref{density})
together with the Maxwell equations.
Thus, the micromagnetic problem  is formulated as
a set of non-linear integro-differential 
equations \cite{Hubert98}.
In many classes of magnetic systems
a strongly pronounced hierarchy of magnetic 
interaction \textit{scales} allows
to reduce the micromagnetic
problem to a set of  auxiliary 
simplified problems \cite{Hubert98,UFN88}.
The procedure includes:
(i) the calculation of spatially homogeneous
equilibrium states by minimizing  energy
\begin{equation}
w_0(\mathbf M) =- {\mathbf M\cdot \mathbf{H}}\ 
+ w_a(\mathbf M)
\label{density1}
\end{equation}
in an (\textit{internal}) magnetic field $\mathbf{H}$ 
for fixed values of the material parameters in Eq.~(\ref{density1}).
The solutions of (i) are used to construct magnetic phase
diagrams in components of the external magnetic field
(ii) and to calculate the equilibrium parameters of multidomain
patterns and the structure of domain walls (iii).

In the rest of the paper, we apply
this program to the model given by Eqs.~(\ref{density})
and (\ref{anisotropy1}).
In order to make transparent the representation of internal
homogeneous states and the phase diagrams, we restrict 
our discussion to the case of co-planar arrangements
of easy axes and applied fields. 
Generalizations of this model are discussed to the
end of the next section.

\section{Reorientation transitions and metastable states}

In many cases of practical interest, 
the direction of the uniaxial anisotropy
$\mathbf{a}$  lies in the plane spanned
by two of the cubic axes $\mathbf{n}_j$ (see Eq. (\ref{anisotropy1})).
To be specific we define this plane as $xOz$ plane assuming that
$z$ is directed along $\mathbf{a}$.
In this case energy $w_0$ from Eq.~(\ref{density1}) 
can be written as a function
of the angle $\theta$ between $\mathbf{M}$ and $\mathbf{a}$. 
Introducing the reduced energy $\Phi(\theta) = w_0/K_c +1/8$ one obtains
\begin{eqnarray}
\Phi (\theta)= -\frac{1}{8}  \cos 4(\theta-\alpha) -\varkappa\,\cos ^2\theta
-h\,\cos (\theta-\psi) \,,
\label{potential} 
\end{eqnarray}
where
\begin{eqnarray}
\varkappa=K_u/K_c, \quad 
h = H /H_c, \quad
H_c = K_c M_0\,,
\label{kappa} 
\end{eqnarray}
$\alpha$ is the angle between the unaxial $\bf a$ 
and cubic $\mathbf{n}_1$ axis, 
the angle $\psi$ defines the deviation of the
magnetic field $\mathbf{H}$ from the easy axis $\mathbf{a}$ in the $xOz$-plane;
correspondingly,  $h_z = h \cos \psi$ is the reduced field
component along the uniaxial easy direction, and
$h_x = h \sin \psi$ is the perpendicular component.

Energy (\ref{potential}) is a function of variable $\theta$
and includes four material (control) parameters, namely,
angle $\alpha$, ratio $\varkappa$ and 
reduced magnetic field components, $h_x$, $h_z$.
Model (\ref{potential}) has been introduced
in 1964 by Torok et al. 
\cite{Torok64} for ferromagnetic films with
misorientated uniaxial and biaxial easy
magnetization directions.
Previous investigations of (\ref{potential})
have been restricted to limiting 
cases of $\alpha =0$ and $\alpha = \pi/4$
and were mostly concentrated on investigations
of coherent rotation processes 
(Stoner-Wohlfarth regime)
(see Ref. \onlinecite{Stoner} and bibliography in Ref. \onlinecite{Hubert98}).
Within this approach switching processes
are identified with the boundaries of
the metastable states (critical astroids).
In this section we give a comprehensive analysis of
model (\ref{potential}) in the full range of the
control parameters ($\alpha, \varkappa, h, \psi$).
In particular, we show that the analysis of the metastable
states only is not sufficient for the understanding of 
magnetization reversal in nanosystems with
competing anisotropies.
The peculiar evolution of
the potential profile (\ref{potential})
under the influence of the
applied field and specific reorientation effects
are found to be crucial for the magnetization 
processes in this class of magnetic materials.

The stationary solutions with the equilibrium values of $\theta$
are derived from the 
equation $\Phi_{\theta} = 0$:
\begin{equation}
h \sin  (\theta -\psi)   =-
\frac{1}{2}\sin  4 (\theta -\alpha) -\varkappa\sin 2 \theta
\label{diff1}
\end{equation}
(Here we introduce a common notation for derivatives
$ f(x)_{x\times k} \equiv
d^k f/dx^k$).
The equation for the lability lines of the solutions,
$\Phi_{\theta \theta} = 0$, reads
\begin{equation}
h \cos  (\theta -\psi)   =-
2 \cos  4( \theta-\alpha)  -2 \varkappa\cos 2 \theta
\label{diff2}
\end{equation}
and determines the stability boundaries 
of the solutions together with Eq.~(\ref{diff1}).
Critical points of the transitions 
are determined by the set of equations
 $\Phi_{\theta\times k} = 0$ ,
 $k = 1,2,3$.
The degeneracy of the solutions, $\theta_i$ with $i = 1...L$,
that provide global energy minima in the system,
\begin{equation}
\Phi(\theta_1)= \Phi(\theta_2)=...=\Phi(\theta_L),
\label{transition1}
\end{equation}
determines the coexistence regions 
between $L$ magnetic phases 
and the conditions for first-order transitions 
in the magnetic phase diagrams.
It is convenient to present the solutions
$\theta (h_x, h_z, \varkappa, \alpha)$  and critical
regions for (\ref{potential})
in a set of phase planes $( h_x, h_z)$ 
parametrized by  the factors $\alpha$ and $\varkappa$.
We start our analysis from the limiting  cases $\alpha=0$
and $\pi/4$,
and after that highlight main features 
of the general model (\ref{potential}).

For symmetric cases with the easy-axes orientations
along one the cubic axis ($\alpha = 0$) and along the
diagonals between them ($\alpha = \pi/4$) the 
Eq.~(\ref{potential}) can be written as 
\begin{eqnarray}
\Phi (\theta)=  \mp \frac{1}{8} \cos (4\theta) -\varkappa\,\cos ^2\theta
-h\,\cos (\theta-\psi)\,, 
\label{potential2} 
\end{eqnarray}
with ``$-$'' for $\alpha =0$ and ``$+$'' for $\alpha =\pi/4$.
Generally model (\ref{potential2}) describes magnetic 
states in a planar ferromagnet with competing
uniaxial (second-order) and biaxial (fourth-order)
magnetic anisotropy. 
The model has been applied for many
bulk and nanoscale magnetic systems,
including reorientation effects
in rare-earth orthoferrites \cite{Belov76,UFN88},
several classes of intermetallic compounds \cite{Melville76},
first-order magnetization 
processes in high-anisotropy materials \cite{Asti80},
and for magnetic nanolayers with surface/interface-induced
magnetic anisotropy 
\cite{Johnson96,Broeder91,Millev98,Oepen00,Lai01,JMMM05,FTT06}.
The model from Eq.~(\ref{potential2}) has also proved to
be valid for diluted magnetic semiconductors 
as a novel class of magnetic materials
\cite{Macdonald05,Sawicki04,Dietl04,JMMM07}.

\begin{figure*}[thb]
\includegraphics[width=15cm,keepaspectratio]{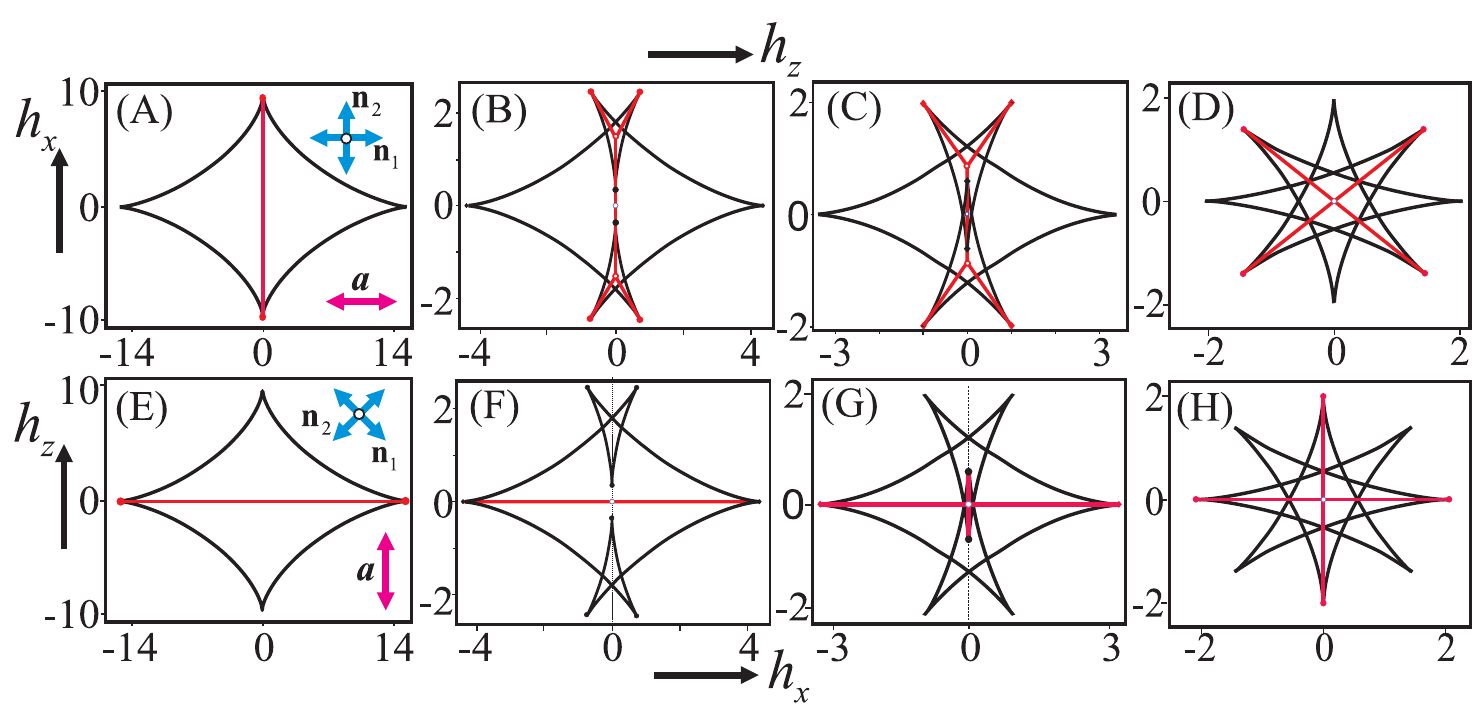}
\caption{
\label{astroids1}
(Color online) The phase diagrams of magnetic states in components of
the internal magnetic field ($h_x$, $h_z$) and different
values of the parameter $\varkappa$ for the ratio of
uniaxial and cubic anisotropy (Eq.\ref{kappa}).
The upper panel (A)-(D) is for systems with $\alpha = 0$,
the bottom panel (E)-(H) is for systems with $\alpha = \pi/4$. 
Two-headed vectors show orientations of 
the uniaxial axis $\mathbf{a}$ and the cubic $\mathbf{n}_i$ axes.
The plots present the topologically different types of
phase diagrams:
(A), (E)  $\varkappa > 5$,
(B), (F)  $5 >\varkappa > 1$,
(C), (G)  $1 >\varkappa > 0$,
(D), (H)  $\varkappa =0 $.
Black lines indicate stability limits of metastable
states.
Red lines give the first-order transitions between
different magnetic phases (see text for details).
}
\end{figure*}

The invariance of the potential Eq.~(\ref{potential2})
under the transformation
\begin{eqnarray}
\varkappa \rightarrow  - \varkappa, \quad
\theta \rightarrow  \theta + \pi/2, \quad
\psi \rightarrow  \psi + \pi/2,
\label{trans} 
\end{eqnarray}
means that the cases with different 
sign of $\varkappa$ transform into each
other by rotation of the reference
system through an angle $\pi/2$.
This invariance allows one to restrict the 
analysis to positive values of $\varkappa$.
The analysis of (\ref{potential2})
yields four topologically different 
types of $(h_x, h_z)$-phase diagrams 
depending on values  $\varkappa > 0$,
as shown in Fig. \ref{astroids1}.

\begin{figure*}[thb]
\includegraphics[width=15cm,keepaspectratio]{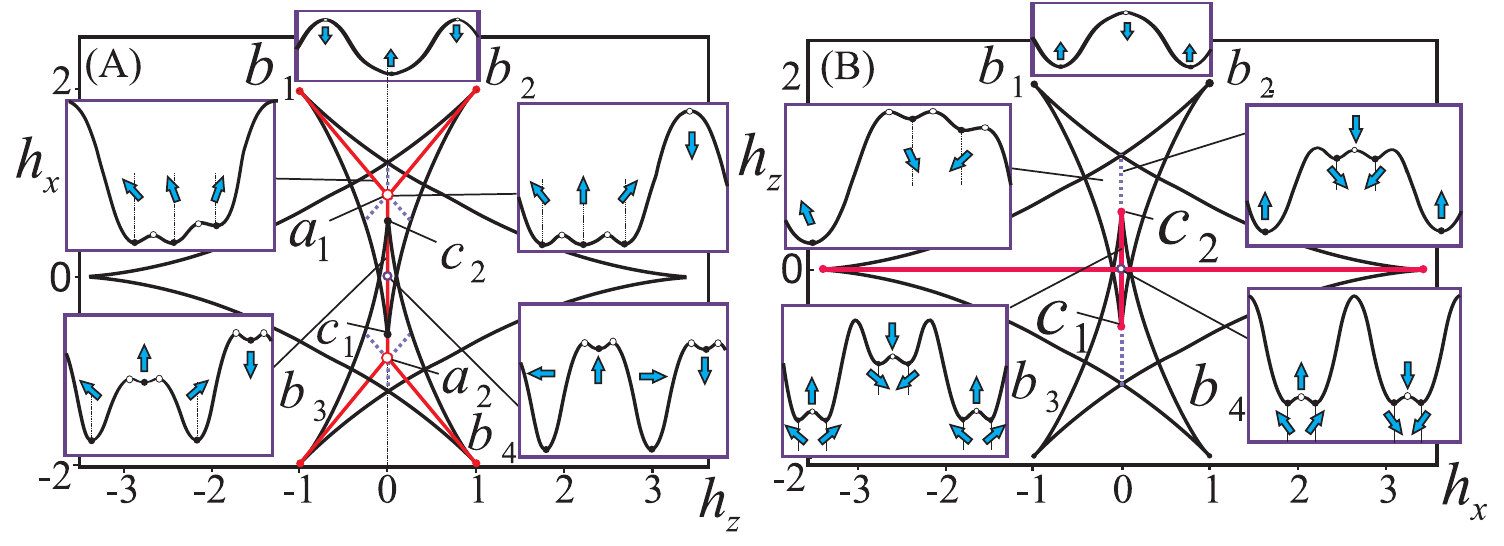}
\caption{
\label{astroids2}
The ($h_x$, $h_z$) phase diagrams 
for $\varkappa = 0.7$: $\alpha = 0$ (A),
$\alpha = \pi/4$ (B).
Potential profiles $\Phi (\theta)$
are sketched for various points 
in the phase diagrams to illustrate the evolution 
of the magnetic states in both models.
}
\end{figure*}

Under transformation (\ref{trans})
the equations of equilibrium (\ref{diff1}), (\ref{diff2}) for the 
potential (\ref{potential2}) with $\alpha =0$ are converted into 
those equations for $\alpha = \pi/4$.
Thus, for the same values of $\varkappa$
the lability lines for both cases transform
into each other by a rotation through $\pi/2$ (Fig. 1). 

For $\varkappa > 5$ the lability lines have a
similar shape as the \textit{Stoner-Wohlfarth}
astroid \cite{Stoner}.
In the limit of large $\varkappa$, the lability line 
asymptotically coincides with this astroid for simple uniaxial ferromagnets.
As $\varkappa$ decreases from 5 to zero the lability
lines transform into an eight-cusp line with the shape of 
the classical wind rose.
At the parameter value $\varkappa = 5$,
a bifurcation takes place by the
formation of so-called ``swallow tails''
\cite{FTT06}
in one pair of opposite cusps (Fig. 1(B), (F)).
In the interval $ 5 > \varkappa > 1$, the swallow
tails gradually widen, and at $\varkappa =1$
the lower cusp points reach the horizontal axis.
In the interval $ 1 > \varkappa > 0$ the phase diagrams
have regions  with  overlapping swallow tails
(Fig. 1(C), (G)).
Finally at $\varkappa =0$, 
that is for zero uniaxial anisotropy,
the lability line transforms into the wind rose with eight corners.
For the two special orientations of the uniaxial anisotropy
in Eq.~(\ref{potential2}) with $\alpha = 0$ and
$\alpha = \pi/4$ 
the lability lines are identical.
Still, the phase diagrams are fundamentally
different as they pertain to different 
magnetic states and different coexistence regions
of metastable magnetic states.
%
%
Depending on the control parameters,
there are regions with  $L=2, 3,$ or $4$ degenerate states,
and, consequently, first-order transitions involving 
\textit{two}, \textit{three} and \textit{four} phases.

For $\varkappa > 5$ the first-order lines
between \textit{two} magnetic states
are segments of straight lines 
connecting opposite cusp points
$h_{z} = 0$, ($|h_{x}| \leq |h_{x}^{c}|=2(\varkappa\mp1)$ 
(Fig. \ref{astroids1}(A), (E)) \cite{FTT06}.
At zero field the transitions occur between
antiparallel magnetic states 
$\theta_1 = 0$ and  $ \theta_2 = \pi$.
For finite transversal magnetic fields
$|h_x| < h_{x}^{c}$ the solutions for
coexisting phases are determined from
Eq.~(\ref{diff1}) with $h_z =0$ 
and $ \alpha =0 \,(\pi/4)$,
\begin{eqnarray}
\sin^3\theta-\frac{1\pm\varkappa}{2} 
\sin\theta\pm\frac{h_x}{4}=0\,.
\label{cub}
\end{eqnarray}
These solutions describe a gradual decrease
of the magnetization component $m_z$.
In the endpoints of the first-order
transitions $h_x= \pm h_{x}^{c}$
the magnetization vectors in both
phases are perpendicular to the easy direction.

For $ 5 >\varkappa\,> 0$ the evolutions
of the magnetic states within the 
swallow tails are different 
for the two models (see
potential profiles 
in Fig. \ref{astroids2}).
For the model with $\alpha =0$, 
the potential wells corresponding  
to the global energy minima are 
swapped within the swallow tails (Fig. \ref{astroids2}(A)).
Hence, different canted states become degenerate in equilibrium 
along lines of first-order transitions (lines $a_1b_1$, $a_1b_2$,
$a_2b_3$, $a_2b_4$ in Fig.~\ref{astroids2}(A)).
These lines meet the transition line
$a_1a_2$ between symmetric phases,
$\theta_1$, $\theta_2 = \pi - \theta_1$,
in the points $a_1$ and $a_2$ 
where \textit{three} phases coexist.
The coordinates of points $a_1$ and $a_2$
are \cite{Melville76,FTT06}
\begin{eqnarray}
\tilde{h}_x = \pm 2 \sin \tilde{\theta}
(\cos 2\tilde{\theta} + \varkappa),
\quad
\tilde{h}_z =0
\label{triple1} 
\end{eqnarray}
and the solutions for the coexisting
phases read
\begin{eqnarray}
\theta_1 = \pm\tilde{\theta},
\quad
\theta_2 =  \pi \mp \tilde{\theta},
\quad
\theta_3 = \pm\pi /2\,,
\label{triple2} 
\end{eqnarray}
where $\sin \tilde{\theta} =
(-1+\sqrt{1 + 3\varkappa})/3$.

For the model with $ \alpha = \pi/4$,
the minimum in the potential is unique 
for the swallow tails
in the parameter range $5 > \varkappa > 1$.
Hence, the transformation of the energy profile 
involves only metastable states
(Fig. \ref{astroids2}(B)).
However, the particular transformation
of the metastable states in these regions
plays an important role in the evolution
of the domain wall profiles (see Sec. VI).
In the interval $1 > \varkappa > 0$
the first order transitions arise
within the region of the overlapping
swallow tails.
The transition line is a segment $c_1 c_2$,
with the points $c_1=(0;2(\varkappa-1))$, $\,c_2=(0;-2(\varkappa-1))$.
Along this line segment
\textit{two} phases coexist with solutions
$\theta_1$ and $\theta_2 = - \theta_1$.
The solutions $\theta_1$ are given by the 
equation
\begin{equation}
\cos^3\theta-\frac{1+\varkappa}{2} \sin\theta-\frac{h_z}{4}=0\,,
\label{cub2}
\end{equation}
that can be derived from Eq.~(\ref{diff1}).
The first-order transition line from $c_1$ to $c_2$ crosses 
the other transition line along $(h_x;0)$ in the origin.
Hence, in this point \textit{four} magnetic phases coexist
with
\begin{equation}
\theta_1=\frac{1}{2}\arccos\varkappa,
\theta_2=-\theta_1,
\theta_3=\pi-\theta_1,
\theta_4=\pi+\theta_1\,.
\label{four}
\end{equation}

For $ \varkappa = 0$ (zero unaxial anisotropy) both potentials
(\ref{potential2})
are converted into the model of a cubic ferromagnet.
The corresponding phase diagrams
(Fig. \ref{astroids1}(D) and (H))
are identical and include two
lines of first-order phase transitions between
symmetric states.
In the origin four degenerate phases
with magnetization along the cubic easy
axes $\mathbf{n}_1$ and $\mathbf{n}_2$ coexist.


\begin{figure}[thb]
\includegraphics[width=8cm,keepaspectratio]{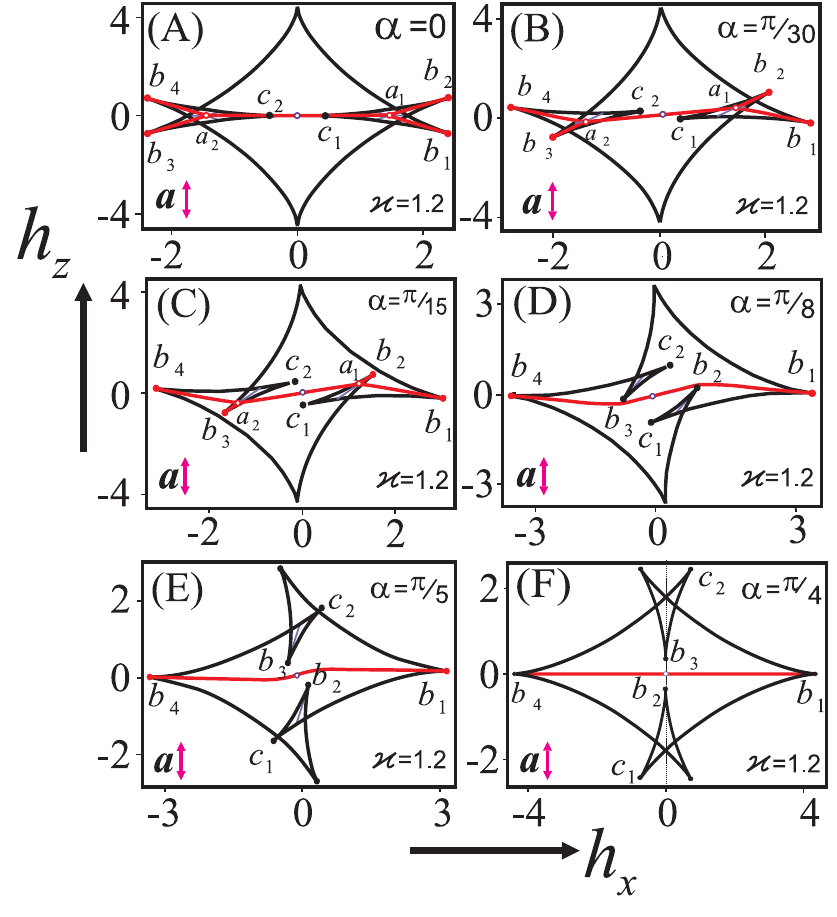}
\caption{
\label{astroids3}
($h_x$, $h_z$) phase diagrams for
$\varkappa = 1.2$ and different 
values of $\alpha$ demonstrate the
transformation between the two symmetric
cases with $\alpha =0$ (A) and 
$\alpha = \pi/4$ (F).
}
\end{figure}

\begin{figure}[thb]
\includegraphics[width=8cm,keepaspectratio]{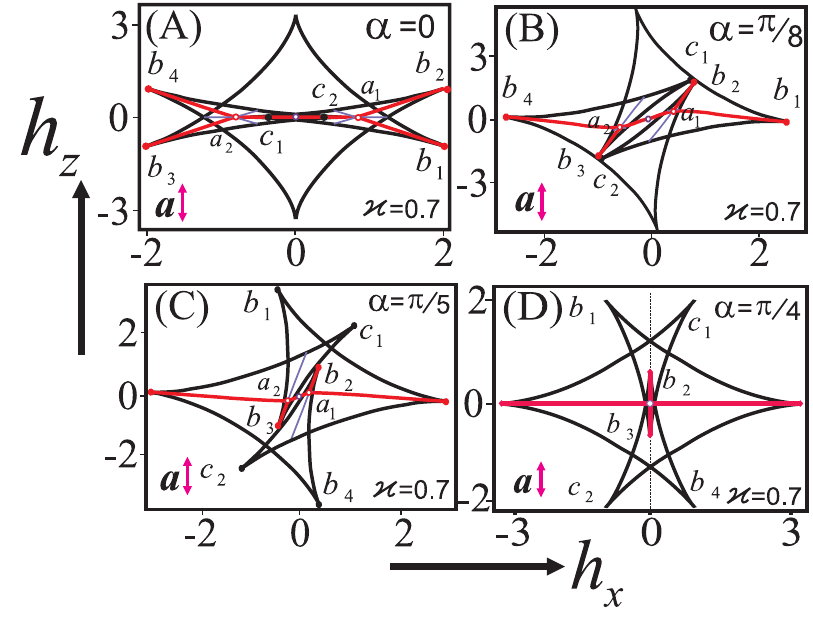}
\caption{
\label{astroids4}
($h_x$, $h_z$) phase diagrams 
similar to that in the previous
figure but for $\varkappa = 0.7$. 
}
\end{figure}

In the general case with a misalignment 
between uniaxial and cubic easy axes given by the parameter $\alpha$
the potential $\Phi$ (Eq.~(\ref{potential})) 
is a periodic function of $\alpha$ with periodicity $\pi/2$.
Thus an analysis in the range $0 \leq \alpha \leq \pi/4$ covers
all physically different states.
Here we describe the evolution of the
$(h_x, h_z)$ diagrams when $\alpha$ varies
from zero to $\pi/4$.
The sets of diagrams in Figs.~\ref{astroids3} 
and \ref{astroids4} show the transformation
of the transition and lability lines.
The case with nonoverlapping swallow tails
for the parameter range $5> \varkappa >1 $ is
presented in Fig. \ref{astroids3}.
For small $\alpha(\varkappa)$ two lines of the
phase transitions between canted phases
still exist (Fig. \ref{astroids3}(B)).
With increasing $\alpha$ the points 
$a_1$ and $a_2$ for the three-phase coexistence
move to either of the cusp points $b_2$ and $b_3$ 
(Fig. \ref{astroids3}(C))
After these points have merged only two-phase 
transition lines exist in the system 
(line $b_1 b_4$ in Fig. \ref{astroids3}(D), (E)).
Thin (blue) lines in Fig. \ref{astroids3}(B)-(F) indicate
the values of the magnetic fields where two metastable
states have the same energy.
They are not connected with any physical processes
in the system but help to understand the transformation
of the energy profiles.

The phase diagram with overlapping swallow tails 
for the parameter range $1 > \varkappa > 0$
is shown in Fig.~\ref{astroids4}.
In this case  the transition lines between
pairs of canted phases in the limit $\alpha = 0$, Fig.~\ref{astroids4}(A),
gradually transform into straight line segments
for transitions between pairs of the symmetric phases 
in the limit $\alpha = \pi/4$, Fig.~\ref{astroids4}(D).
During this process the points of 
the three-phase coexistence $a_1$ and $a_2$ 
move towards each other (Fig. \ref{astroids4}(B), (C)),
and merge into the point with four-phase coexistence at the origin
for $\alpha = \pi/4$, Fig.~\ref{astroids4}(D).

%
%
The sets of modified astroids in Figs.~\ref{astroids1}
and \ref{astroids3} represent geometrical singularities
studied by a special field in mathematics known as
\textit{catastrophe theory} \cite{Poston78}.
It was found that for rather general form
of potentials there exist only
seven fundamental types of singularities
referred to as \textit{catastrophes} \cite{Poston78}.
Four of them are realized in the lability
lines of Figs.~\ref{astroids1} and \ref{astroids3}.
The astroid lines, where one local minimum
merges with a local maximum, are named \textit{fold catastrophes}.
The edge points where two folds meet have
two minima merging with a maximum.
These singularities are known as \textit{cusp catastrophes}.
By a characteristic discord in the terminology,
the feature known in magnetics 
as ``swallow tails'' as shown in Fig. 1(B), (F)
are called \textit{butterfly catastrophes} in mathematics,
while the triangular regions, as those with the cusps
$c_1$,$b_2$ in Fig. \ref{astroids3}(D),
are called \textit{swallow tail catastrophes}.
The lability lines in Fig. \ref{astroids1}
belong to a family of \textit{hypercycloids}.
In present article we shall adhere to 
the terminology used in micromagnetism.

The transformation of the common
Stoner-Wohlfarth astroid (as 4-cusped hypercycloid-
Fig. \ref{astroids1}(A), (E)), into
the 8-cusped hypercycloid with the wind rose shape
(Fig. \ref{astroids1}(D), (H)) occurs in
many magnetic systems with competing
anisotropies and has been investigated
during the last forty years.
To the best of our knowledge the
8-cusped hypercycloid has been
firstly obtained in Refs. \onlinecite{Torok64,Yelon65}  
(see also the remarks about earlier conference
contributions in Ref. \onlinecite{Torok64}). 
The transformation from the common astroids 
(Fig. \ref{astroids1})
into a lability line with swallow tails, and the
further evolution of these curves to the wind rose
has been obtained in Ref. \onlinecite{Torok64}. 
Torok et al. also demonstrated several diagrams
of lability line for model with
misorientated anisotropy axes
(Eq.(\ref{potential2})).  
In many following papers (see, e.g.
Refs. \onlinecite{Torok65,Mitsek1,Mitsek2}) 
pecularities of lability lines
for the model~(\ref{potential2}) have been 
investigated.
The coordinates of the critical
points for (\ref{potential2})
were calculated in Refs. \onlinecite{Melville76,Chang88A,Chang88B,Kaganov}. 
The solutions for the first-order 
phase transition lines and the coexisting
states have been carried out in Refs. \onlinecite{FTT87,FTT87b} 
(see also Ref. \onlinecite{FTT06})). 
In this paper we have given an exhaustive summary
of model (\ref{potential}).

\section{Magnetic phase diagrams}

\begin{figure*}[thb]
\includegraphics[width=15cm,keepaspectratio]{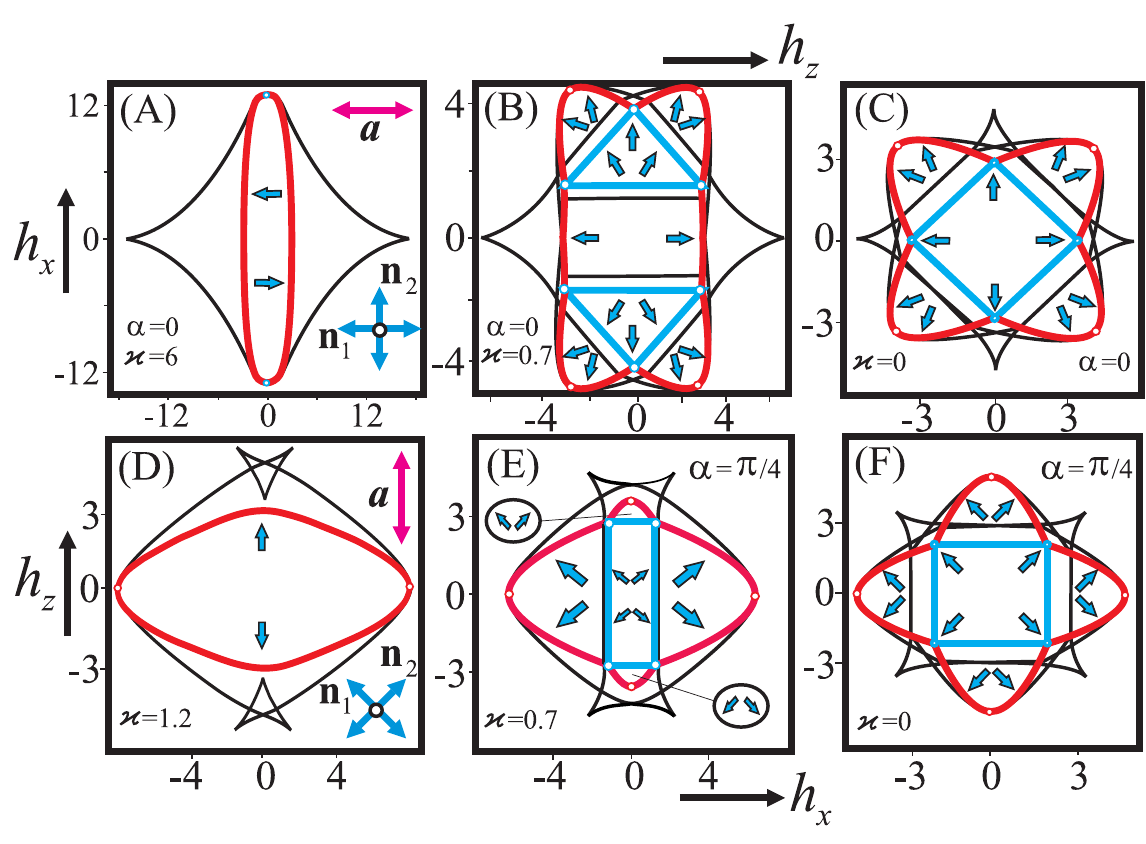}
\caption{
\label{he}
Magnetic phase diagrams in the components
of the external field
$h_x^{(e)}$, $h_z^{(e)}$
for  $\alpha = 0$ ( (A)- (C))
and $\alpha = \pi/4$ ((D)- (F)).
Thick lines limit regions of
three- and four-phase  (blue)
and two-phase (red) multidomain states.
Arrows show magnetic configurations in
the (co-existing) domains.
All calculations have been carried out
for a spherical sample ($N_{ii} =1/3$).
}
\end{figure*}
\begin{figure*}[thb]
\includegraphics[width=15cm,keepaspectratio]{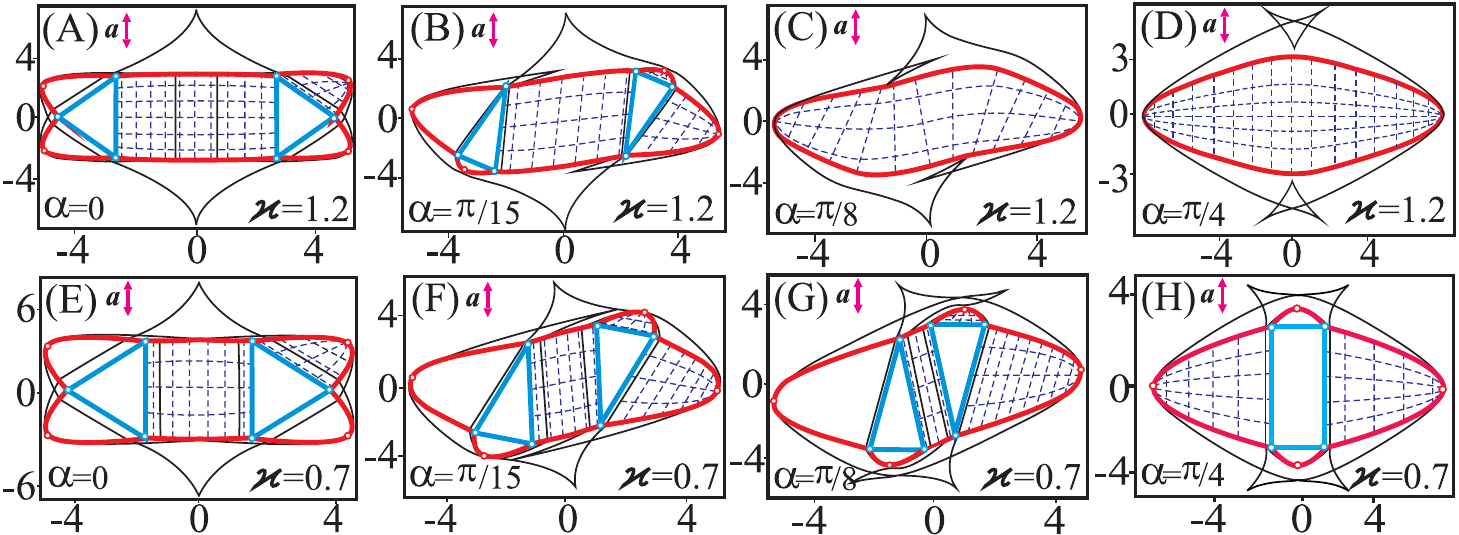}
\caption{
\label{he13}
Magnetic phase diagrams in the components
of the external field
$h_x^{(e)}$, $h_z^{(e)}$
for  $\varkappa=1.2$: (A)- $\alpha=0$; (B)- $\alpha = \pi/15$;
(C)- $\alpha = \pi/8$; (D)- $\alpha = \pi/4$ and $\varkappa=0.7$: 
(E)- $\alpha=0$; (F)- $\alpha = \pi/15$;
(G)- $\alpha = \pi/8$; (H)- $\alpha = \pi/4$. 
Thin dotted lines are the lines of the constant 
internal field and
the constant phase fractions.
}
\end{figure*}

In the previous section the solutions
for possible magnetic states have
been presented as functions of the
internal field.
For ellipsoidal magnets with a homogeneous
magnetization $\mathbf{M}(\mathbf{h})$ 
the equation \cite{Hubert98} 
\begin{equation}
\mathbf{h}^{(e)} = \mathbf{h} 
+ 4 \pi K_c^{-1} \hat{\mathbf{N}}\mathbf{m}(\mathbf{h})
\label{ellipses1}
\end{equation}
establishes the correspondence between magnetic phase
diagrams in terms of the internal field $\mathbf{h}$ and 
those in terms of the external field $\mathbf{h}^{(e)}$
($\hat{\mathbf{N}}$ is the demagnetizing tensor).
For phase diagrams in Figs. \ref{astroids1} and
\ref{astroids3} the phase diagrams
in external magnetic field components are
plotted in Figs. \ref{he} and \ref{he13},
correspondingly.

Thin lines in Figs. \ref{he}, \ref{he13}
define values of the external fields
in which the internal field within 
the stable phases reaches the boundaries
of the metastable region.
The transition lines ($\mathbf{h}_{tr}$) 
in Figs.~\ref{astroids1}, \ref{astroids3}
transform into the areas bounded by thick (red)
lines in Figs. \ref{he}, \ref{he13}. 
These areas define maximum possible regions
where thermodynamically stable multidomain
states of the competing phases can
exist \cite{Hubert98,UFN88,FNT98}.

For $\varkappa > 5$ and
$\alpha =0$ 
the ($h_x, h_z$) diagram in Fig.~\ref{astroids1}(A)
converts into that in Fig.~\ref{he}(A).
The transition line in Fig.~\ref{astroids1}(A)
transforms into an area of a two-phase multidomain state 
that is bounded by an ellipse (red line). 
For $|\varkappa| < 5$ and $\alpha =0$
the diagrams in the terms of external-field components
become rather complicated: Fig. \ref{he}(B)
is obtained by mapping the phase diagram in
Fig.~\ref{astroids1}(B),
and Fig.~\ref{he}(C) by mapping 
Fig.~\ref{astroids1}(D).
The lines of the two-phase transitions in
Fig. \ref{astroids1}(B), (C)  convert
into areas of two-phase domain structures (DS);
the points  of three-phase coexistence
(\ref{triple1}) ``swell'' into the triangular
regions of three-phase domain structure;
and the point (0,0) in Fig.~\ref{astroids1}(D),
where four phase $\theta_j=\pi j/4$ coexist, 
transforms into a rectangular area 
with a four-phase multidomain state.

The phase diagrams in Fig.~\ref{astroids1}(F), (G), (H)
for systems with $\alpha = \pi/4$
are mapped into the phase diagrams in Fig. ~\ref{he}(D), (E), (F),
correspondingly.
For $\varkappa > 1$ the phase diagram in
Fig.~\ref{he}(D) includes one region
of two-phase multidomain states
and two swallowtail pockets in the metastable region.
For $1 > \varkappa >0$ $\mathbf{h}^{(e)}$-phase
diagrams include a rectangular area, where 
four-phase domain states are stable 
with spin configurations in the domains 
described by Eq.~(\ref{four}).
Adjacent to this area, there are four regions 
in the phase diagrams with two-phase
multidomain states (Fig. \ref{he}(E)).
Finally the $\mathbf{h}^{(e)}$-
phase diagram for $\varkappa =0$ in Fig.~\ref{he}(F) becomes
identical to that in Fig.~\ref{he}(C).

For  $\alpha$ varying from zero
to $\pi/4$ the phase diagrams are plotted in the case $\varkappa=1.2$ (Fig. \ref{he13}(A)-(D)) 
and $\varkappa=0.7$ (Fig. \ref{he13}(E)-(H)) and reflect the complex transformation
of the regions of the DS existence. 
For $\varkappa=1.2$ the triangular areas of three phase DS
and the regions of two phase DS deform (Fig. \ref{he13}(B)) 
and then disappear at all (Fig. \ref{he13}(C)) 
leading to a phase diagram with one distorted ellipse of  two phase DS.
The phase diagram for $\varkappa=0.7$ demonstrate another kind of transformation.
Now, the regions of three-phase DS do not disappear (Fig. \ref{he13}(F), (G)) but, on the contrary, 
join to form a rectangular area of four phase DS (Fig. \ref{he13}(H)).
Here in Fig. \ref{he13}, the thin dashed lines denote the lines of constant internal field 
and constant phase fractions (in regions of stable two-phase DS).
When an external magnetic field is varied
following these lines, then either domain walls
are displaced or the magnetization rotates in each
domain, respectively.

In the next section we apply the diagrams of
solutions in Figs. \ref{astroids1},\ref{astroids2},\ref{astroids3},\ref{astroids4}
and the phase diagrams
in Figs. \ref{he},\ref{he13}
to analyze the magnetization 
processes in nanosystems with competing
anisotropies.
We emphasize that those diagrams describe
two limiting cases of ideally
hard and ideally soft magnetic behavior.
In ideally hard magnetic materials 
magnetization processes occur via evolution
of metastable states.
Magnetic phases exist everywhere in
their stability regions up to their 
boundaries (\textit{Stoner-Wohlfarth
regime}).
The corresponding magnetization curves
(dotted lines in Fig. \ref{curves})
are characterized by the widest possible 
hysteresis cycles \cite{Hubert98},
and single domain states are realized in 
these systems.
In the opposite case of ideally
soft magnetic materials the magnetization
reversal occurs via the evolution of
\textit{thermodynamically} stable
states.
Such \textit{anhysteretic} processes involve
the formation of multidomain patterns.
These spatially inhomogeneous states are
composed of domains formed by the competing phases
of the magnetic-field induced first-order
transition \cite{Hubert98,UFN88}.
Extended regions
of multidomain states
have been observed
during reorientation processes in
several groups of bulk magnetic systems 
(e. g. orthoferrites
and easy-axis antiferromagnets
\cite{UFN88,PRB07}).
For these magnetically 
soft, low anisotropy systems the
multidomain states are described by the
phase theory equations \cite{Hubert98,UFN88}.
The phase theory approximation stricly
is valid only if the
characteristic sizes of the sample 
are much larger than the sizes of
domains, and transitional regions
between domains are localized
into narrow domain walls \cite{Hubert98}.
Both these requiments
are met only in ideally soft, massive
magnetic samples.
Thus, in soft magnetic materials the magnetization
processes are mainly determined by occurrence of 
the first-order phase transitions and the evolution
of the magnetic states  in the coexistence phases
during these transitions (solid lines in Fig. \ref{curves}).
The magnetization processes in real materials 
are intermediate between these two limiting cases and
include both evolution of the metastable
and multidomain states.
In magnetic nanolayers domain sizes
usually exceed the layer thickness.
In magnetic nanoparticles 
only few domain walls are observed,
and in sufficiently small particles
multidomain states are completely 
blocked.
On the other hand, coercivity
of the magnetic nanosystems prevents 
the formation of the equilibrium states
and causes hysteretic magnetization 
reversal.

\section{Comparison with experiment}

The phase diagrams of solutions
in Figs. \ref{astroids1}-\ref{astroids4}
can be applied for explanation of 
magnetization processes in many nanomagnetic systems
with competing anisotropies,
for example, in thin films of diluted magnetic semiconductors (DMS),
in ferromagnetic(FM)/antiferromagnetic(AFM) bilayers \cite{Wang05,Lai01},
Heusler alloys \cite{Yang02}, and/or nanoparticles \cite{APL07,JMMM05}.
First, we consider the phase diagrams with symmetric
arrangement of easy uniaxial and cubic anisotropy axes
(Fig. \ref{astroids1},\ref{astroids2}),
as applied for nanolayers of DMS.
And then we give examples of systems with different values of
angle $\alpha$ between anisotropy axes.

Layers of diluted magnetic semiconductors represent a
new class of materials with 
a strongly pronounced
competing character of the magnetic anisotropy.
In existing (Ga, Mn) As nanolayers
the ratio $\varkappa$ of uniaxial and cubic anisotropy varies in a broad range
and is controlled by strains, temperature and hole concentration
\cite{Macdonald05,Sawicki04,Dietl04,Sawicki004,Sawicki05,Sawicki06,Wang05}.
The  magnetization
processes in Ga$_{1-x}$Mn$_x$As thin films grown
by molecular beam epitaxy on GaAs(001) substrates
are described by the diagrams of solutions 
for highly symmetric geometry, $\alpha=0;\pi/4$
(Fig. \ref{astroids1}).
The in-plane magnetization reversals in these systems
are determined by the competition of cubic anisotropy
with easy axes $<100>$ and uniaxial
anisotropy favouring the directions of $<110>$ type.
Thus, the solutions of (\ref{potential}) for $\alpha=\pi/4$
are applicable in this case (Fig. \ref{astroids1}(E)-(H)).
The main features of in-plane magnetization processes in such
layers are summarized in Fig. \ref{astroids2}(B)
and were experimentally investigated 
in a number of works \cite{Sawicki004,Liu03,Welp03,Moore03}.

\begin{figure*}[thb]
\includegraphics[width=17cm,keepaspectratio]{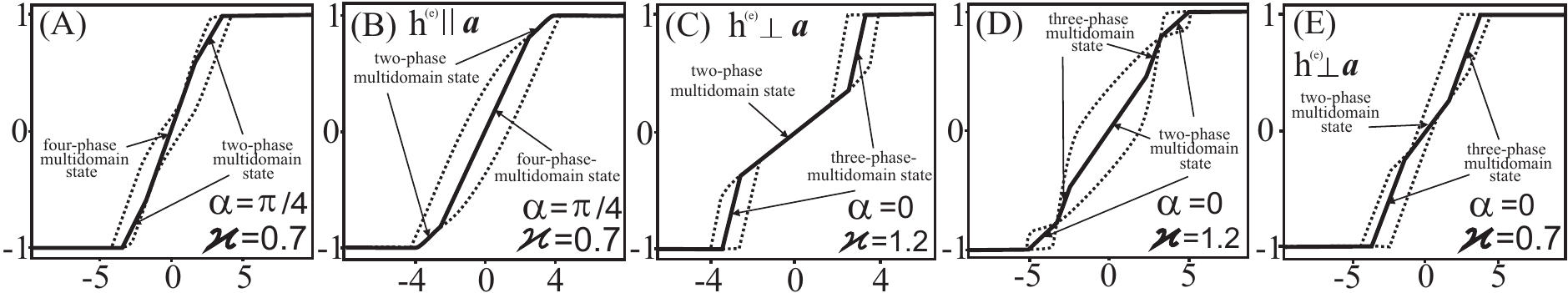}
\caption{
\label{curves}
Magnetization curves (schematically)
for systems with two- and four-phase multidomain states (A),(B)
and for those with two- and three-phase multidomain states(C)-(E).  
}
\end{figure*}

In Ref. \onlinecite{Welp03}
the 300nm thick $\mathrm{Ga_{1-x}Mn_xAs (x=0.03)}$
samples  were studied combining 
direct imaging of magnetic domains and SQUID
magnetometry.
At temperatures above 30K the samples
exhibit the uniaxial anisotropy with
easy axis along [110], whereas for
temperatures below 30K the
magnetization vector deviates from
this direction indicating the dominance
of the fourfold symmetry.
According to our phenomenology
these magnetic films followed the temperature transition
from the phase diagram in Fig. \ref{astroids1}(F)
for dominating uniaxial anisotropy ($\varkappa>1$) to
that in Fig. \ref{astroids1}(G)
with competing character of anisotropy ($\varkappa<1$).
The angle between the magnetization and
the axis [110]
is determined by the Eq. (\ref{four}).
For T=15K (the ratio $\varkappa=0.42$
was obtained from fits of 
the hard-axis magnetization curves
and calculating the easy-axis orientation
from Eq.(\ref{potential2})) 
this angle is 32$^\mathrm{o}$
which agrees with experimental results.
The magnetization processes for high temperature
($\varkappa>1$, Fig. \ref{astroids1}(F), \ref{he}(D))
proceed through the nucleation and expansion
of domains with two orientations of
the magnetization vector.
In fields applied along the easy axis [110]
the evolution of the DS is accompanied 
only by the 180$^\mathrm{o}$ domain wall movement
(Fig. \ref{he13}(D)),
and the metastable states in swallow tails
(see energy profiles in Fig. \ref{astroids2}(B)) 
can be considered as the nuclei of domains.
During the magnetization processes
along [100] axis the domains not only
nucleate and expand but the magnetization
rotates inside each domain of various
phases (Fig. \ref{he13}(D)). 
The magnetization reversal for low temperature
($\varkappa=0.42$)
along one of the cubic easy axes
proceeds in three stages\cite{Welp03} through  the formation
of intermediate domains (Fig. \ref{astroids1}(G), \ref{he}(E)).
In the first (and the last) stage a transversely magnetized domain
nucleates indicating the entering into 
the area of two-phase DS in Fig. \ref{he}(E).
With field increasing the completely reversed
domains nucleate and propagate rapidly through
the sample indicating the beginning of the
area with four phase DS in Fig. \ref{he}(E) \cite{Welp03}.
It is remarkable that the one stage switching processes 
are also possible and are accompanied by
the transformation of four phase domain structure
for some directions of magnetic field
(blue open points in Fig. \ref{he}(E)).
The magnetization curves in Fig. \ref{curves} (A), (B)
are typical for the in-plane geometry displaying biaxial character
of the anisotropy. 
The successive switching of the magnetization 
in Fig. \ref{curves}(A) are caused by
the redistribution of the metastable minima in the
energy profiles for varying magnetic field (Fig. \ref{astroids2}(B)).
The hysteresis loops of such a type
are more pronounced
for purely cubic anisotropy ($\varkappa=0$)
and were observed  for 603 nm-$\mathrm{Ga_{0.957}Mn_{0.043}As}$
films \cite{Moore03}.
If the applied magnetic field makes the angle with the [100] axis in the 
range $(0;\pi/4)$, and initially spins are aligned
along [$\overline{1}00$], then the first incoherent reversal 
is related to the appearance of domains with [010]
magnetization, whereas the second jump is due to
the [100] domain (Fig. \ref{astroids1}(D)).

In Ref. \onlinecite{Pappert07} the character 
of in-plane magnetic anisotropy has been determined
by means of transport measurements.
All layers were patterned into 40-60$\mu$m wide
Hall bar structures, and a strong anisotropic magnetoresistance
effect provides a very convenient method to study the
anisotropy at fixed temperature.
The resistance polar plots of transport measurements
for prevalent biaxial anisotropy \cite{Pappert07}
look similar to the phase diagram in Fig. \ref{he}(C).
The [110] uniaxial anisotropy
leads to the narrowing and subsequent disappearance
of the four phase DS area (Fig. \ref{he}(D), (E), (F)).
As well, it was shown 
that an additional uniaxial anisotropy 
with [010] easy axis is present in the system.
This anisotropy results
in the formation of a two phase DS region
splitting the rectangle with four phases
(Figs. \ref{he}(B), (C)).
 
The solutions with $\alpha=0$ (Fig. \ref{astroids1}(A)-(D)).
are realized 
for out-of-plane magnetic field and in-plane
orientation of the magnetization \cite{Titova05,Liu03,Liu05}.
The lability lines of phase diagrams for $\alpha=0$
(Figs. \ref{astroids1}(B), (F))
are similar to those for $\alpha=\pi/4$
but the magnetization processes are quite different.
The triple point with
three coexisting phases inside the swallow tail (Fig. \ref{astroids1}(B))
has a crucial influence on the magnetization reversal
and is the reason of specific double shifted hysteresis
loops (Fig. \ref{curves}(C)) observed in many works \cite{Titova05,Sawicki004,Sawicki06}.

In Ref. \onlinecite{Titova05}  
$\mathrm{Ga_{1-x}Mn_xAs}$ films grown on hybrid
ZnSe/GaAs substrates with
a low Mn concentration $(x\approx0.01)$ were
chosen to identify the role of both types
of anisotropies in the magnetic reversal process.
Varying the hole concentration $p$ and
temperature $T$ the ratios $\varkappa$ according
to Figs. \ref{astroids1}(A)-(D) can be swept.
For the hole concentration $p=8.5\cdot10^{19}\mathrm{cm}^{-3}$
the temperature progression results in the succession
of phase diagrams, namely, Fig. \ref{astroids1}(A)
for T=20K, Fig. \ref{astroids1}(B)
for T=7K, Fig. \ref{astroids1}(C)
for T=3K and Fig. \ref{astroids1}(D)
for T=1.5K.
The switching processes for high temperature
exhibit the typical behavior of a specimen
magnetized along the hard direction (Fig. \ref{astroids1}(A)).
In this case, a domain structure exists with
magnetization vector tilted with respect to the
magnetic field.
As the temperature is lowered the triple point
in Fig. \ref{astroids1}(B) denotes the existence
of an additional stable magnetization state 
along the magnetic field direction.
Thus, the subloops of the hysteresis
curves (Fig. \ref{curves}(C)) reflect the
jump of the magnetization into this minimum
accompanied by the three phase domain structure.
The variation of magnetic
field in the region spanned by the swallow tails 
lead to various scenarios of the DS transformation
(see energy profiles in Fig. \ref{astroids2}(A)).
In particular, different cases (shown in Fig. \ref{he}(B) by
the red and blue open points) of the transition
from multidomain states into a single domain state can be realized.
For some directions of the magnetic field
crossing regions with two- and three-phase DS 
one can observe even more complex hysteresis 
loops consisting of three subloops (Fig. \ref{curves}(D)).
In our phenomenology the maximum hysteresis
loops encircle the anhysteretic magnetization curves
with three and two phase DS.
Experimentally that kind of magnetization 
processes was observed in Co$_2$MnSi and Co$_2$MnGe
Heusler alloys \cite{Yang02}.
With the temperature decreasing the subloops in Fig. \ref{curves}(C)
broaden around the two steps of the magnetization,
and a hysteresis loop with only a weak
double shift is observed (at T=1.5K, the experimentally
measured value of $\varkappa\approx{0.26}$ corresponds to phase diagram
Fig.\ref{curves}(E)).
Note, that one should distinguish the hysteresis
loops of such type for the cases with $\alpha=\pi/4$ (Fig. \ref{curves}(A))
and $\alpha=0$ (Fig. \ref{curves}(E))
because the magnetization processes are fundamentally
different.
For (Ga,Mn)As systems with low hole concentration 
($p=3.0\cdot10^{19}\mathrm{cm}^{-3}$)
uniaxial anisotropy is much larger than the cubic anisotropy \cite{Titova05}.
This situation is described by model
(\ref{potential}) with  $\varkappa>5$
(Fig. \ref{astroids1}(A), \ref{he}(A)).
If the magnetic field is applied along the
easy axis of uniaxial anisotropy (Fig. \ref{astroids1}(B),\ref{he}(B))
then the two phase DS transforms
into a single phase state.
As a remnant of the DS, 360$^\mathrm{o}$
domain walls may remain in this state and can 
act as nuclei of new reverse domains when lowering
or changing the external magnetic field.
Experimentally such situations have been studied 
in magnetic field perpendicular to the film and for
out-of-plane magnetization vector \cite{Thevenard06}. 

In FM/AFM bilayers of cubic materials
the intrinsic cubic $<100>$ anisotropy may
compete with the uniaxial anisotropy induced by the
exchange couplings between two layers \cite{Wang05}.
To establish the exchange bias uniaxial anisotropy, the bilayer film
is cooled in an in-plane magnetic field which determines
the easy axis of induced anisotropy. 
In Ref. \onlinecite{Wang05} the magnetization reversal has been studied
in an exchange-biased CaMnO$_3$/La$_{0.67}$Sr$_{0.33}$MnO$_3$ bilayer film
grown on vicinal SrTiO$_3$ $<100>$ with
the angle between cubic and uniaxial easy axes being $\alpha=30^\mathrm{o}$.
With temperature decreasing the magnetic films followed
the transition from the phase diagram of solutions in Figs. \ref{astroids3}(E) (T=160K)
to that in Fig. \ref{astroids4}(C) (T=5K).
The magnetization processes for high temperatures involve
only the redistribution of a two phase DS (Fig. \ref{he13}(C)).
The metastable states inside each swallow tail lead
only to the slight deformation of energy profiles and 
do not influence significantly the magnetization reversal
(Fig. \ref{astroids3}(E)).
For low temperatures those metastable states become
stable (Fig. \ref{astroids4}(C)) and alternatively
a domain state can be realized.
The three and two phase DS (Fig. \ref{he13}(F),(G))
in the system result in complex  hysteresis loops with
a hint of a double shift
(Fig. \ref{curves}(E)) \cite{Wang05}.

Double shifted magnetization curves
with strongly pronounced subloops (Fig. \ref{curves}(C))
and the astroid with swallow tails (Fig. \ref{astroids3}(A))
were observed in Ref. \onlinecite{Lai01} 
for metallic multilayersamples with the structure Si(100)/Cu(15nm)/
Ni$_{80}$Fe$_{20}$(35nm)/NiMn(50nm)/Co(tnm)/Pd(15nm)
grown by an e-beam evaporation system.
The thickness was varied
between 5 nm and 25 nm.
It was shown that the double-shifted
hysteresis loops (and parameter $\varkappa$) could be tuned
by several parameters, e.g., the variation of Co film thickness,
and the field-annealed time.

In magnetic single-domain nanoparticles the competition
of the uniaxial anisotropy
due to the enhanced surface interactions and intrinsic 
magnetocrystalline anisotropy stabilizes different
multiple magnetic states in the system with the 
possibility of switching between them
(for details see Ref. \onlinecite{APL07}).
Different geometries of relative easy axes alignment are realized in 
these nanoobjects.
Phase diagrams with "swallow
tails" (Fig. \ref{he13}(C)) for misaligned easy anisotropy
axes have been obtained for Fe–Cu–B
nanoparticles \cite{Duxin00} and Co clusters \cite{Jamet04}.
Astroids with rounded  corners (Fig. \ref{he}(A))
have been observed in fcc-Co \cite{Wernsdorfer02}
and BaFeCoTiO nanoparticles \cite{Wernsdorfer97}.

In general, in many cases of practical interest
the geometry with noncoplanar competing anisotropies 
is observed \cite{Wang05,311A,311B}.
For example, in 50 nm thick Ga$_{0.91}$Mn$_{0.09}$As thin films grown on
(311)A and (311)B substrates\cite{Wang05} the uniaxial anisotropies
with [01$\overline{1}$] (or [$\overline{2}33$])
and [311] easy axes compete with cubic anisotropy of $<100>$ type.
In this case three dimensional phase diagrams,
parametrized by the varios ratios of anisotropy
coefficients and magnetic field components,
have to be constructed instead of 2D diagrams of
solutions.
Even for considered DMS films with (001)
orientation, the out-of-plane magnetization
processes can be considered as coplanar only with some
restrictions.
Indeed, for biaxial in-plane anisotropy and
magnetic field perpendicular to the film
one generally has a non-coplanar arrangement of
the magnetization in domains.
In that case, the phase diagram
in Fig.\ref{he} (B) is only the cross-section
of that more complex 3D phase diagram.
But due to the degeneracy of in-plane cubic [100]
and [010] axes with respect to the magnetic field
the main peculiarities of the switching
processes can be readily explained with
the simple 2D phase diagram (Fig.\ref{astroids1}(B)).
Therefore, the magnetic anisotropy
geometry and magnetic field orientation 
 determine which phase diagrams
of solutions (2D or 3D) is applicable in a
particular case.


\section{Multidomain patterns} 
\label{Walls}

Multidomain patterns have been
observed in a number of systems
with in-plane \cite{Welp03} and
out-of-plane magnetization
\cite{Shono00,Fukumura01,
Pross04,Thevenard06,Shin07,
Thevenard07,Dourlat07,Gourdon07}.
Particularly, isolated domain walls
trapped in micropatterned constrictions
of (Ga,Mn)As films demonstrate a number
of remarkable properties 
\cite{Versluijs01,DW1,DW2,DW3}
and can be used in differerent
nanoelectronic devices (e. g. as
magnetoresistive elements) 
\cite{Versluijs01}.
The fine structure 
of the isolated domain wall
is of prime importance
when different types of domain 
walls are observed \cite{Sugawara07}. 
Here in particular, we demonstrate that
for the considered systems with competing
anisotropies various types of domain walls
exist with large sensitivity of their appearance
on material parameters and external fields.
Using the results of the two previous
sections we calculate the equilibrium 
parameters of isolated planar
domain walls and derive the equilibrium
parameters of stripe domains in system
with out-of-plane magnetization.

\subsection{The structure of domain walls}

For a planar isolated domain wall 
with energy density
$\Phi (\theta)$ (\ref{potential})
the structure is derived by optimization 
of the functional
\begin{equation}
w_{DW}=A \theta_x^2+K_c \Delta \Phi(\theta)
\label{DW}
\end{equation}
with the boundary conditions 
$\theta_x (\pm \infty) = 0$, $\theta(+ \infty) = \theta_1$,
$\theta(- \infty) = \theta_2$
($x$ is a spatial variable across the domain wall),
and
$\Delta \Phi(\theta)= \Phi(\theta)-\Phi_0 $, where
$\Phi_0 = \Phi(\theta_1) = \Phi(\theta_2) = \{\textrm{min} \Phi \}$ 
is the global minimum of the system.
For such a one-dimensional problem (\ref{DW}) 
the Euler equation and the first integral 
can be written as \cite{Hubert98}
\begin{subequations}
\begin{align}
&2x_0^2 \theta_{xx}=\Phi_{\theta},\label{Euler1}\\
\quad
&x_0^2 (\theta_x)^2=\Delta \Phi(\theta) \label{Euler2}
\end{align}
\label{Euler}
\end{subequations}%
where $x_0= \sqrt{A/K_c}$ is a charactristic width
of the domain wall.

The domain wall profiles $\theta(x)$, their energy
and characteristic sizes can be  readily derived by
direct integration.
However, Eqs. (\ref{Euler1}) allow to understand
the main features of such solutions.
The Eq. (\ref{Euler1})
shows that the \textit{inflection} points of the domain
wall profiles $\theta (x)$ correspond 
to \textit{stationary} points of potential (\ref{potential})
($\Phi_{\theta} =0$).
The second equation (\ref{Euler2}) shows that the larger the deviations
of the energy from the minima $\Delta\Phi(\theta)$
the larger the gradients of the profiles, $\theta_x$.
Thus, the nucleation and further evolution of
local minima in the metastable region causes complex
reconstructions of the domain wall profiles.
Transformations of the domain wall profiles
have been earlier observed in easy-axis antiferromagnets
and other magnetic crystals (see examples 
in Ref. \onlinecite{UFN88}). 
Due to the remarkable lability of the potential profiles
(\ref{potential}) (Fig.\ref{astroids2}) this effect is expected
to be strong in the systems with competing
anisotropies.
As an example, 
we derive the parameters of the domain
walls for four-phase domains with the canted states
(\ref{four}) that are realized in the systems 
with $\alpha=\pi/4$, $1> \varkappa > 0$  at zero fields.

\begin{figure*}[thb]
\includegraphics[width=15cm,keepaspectratio]{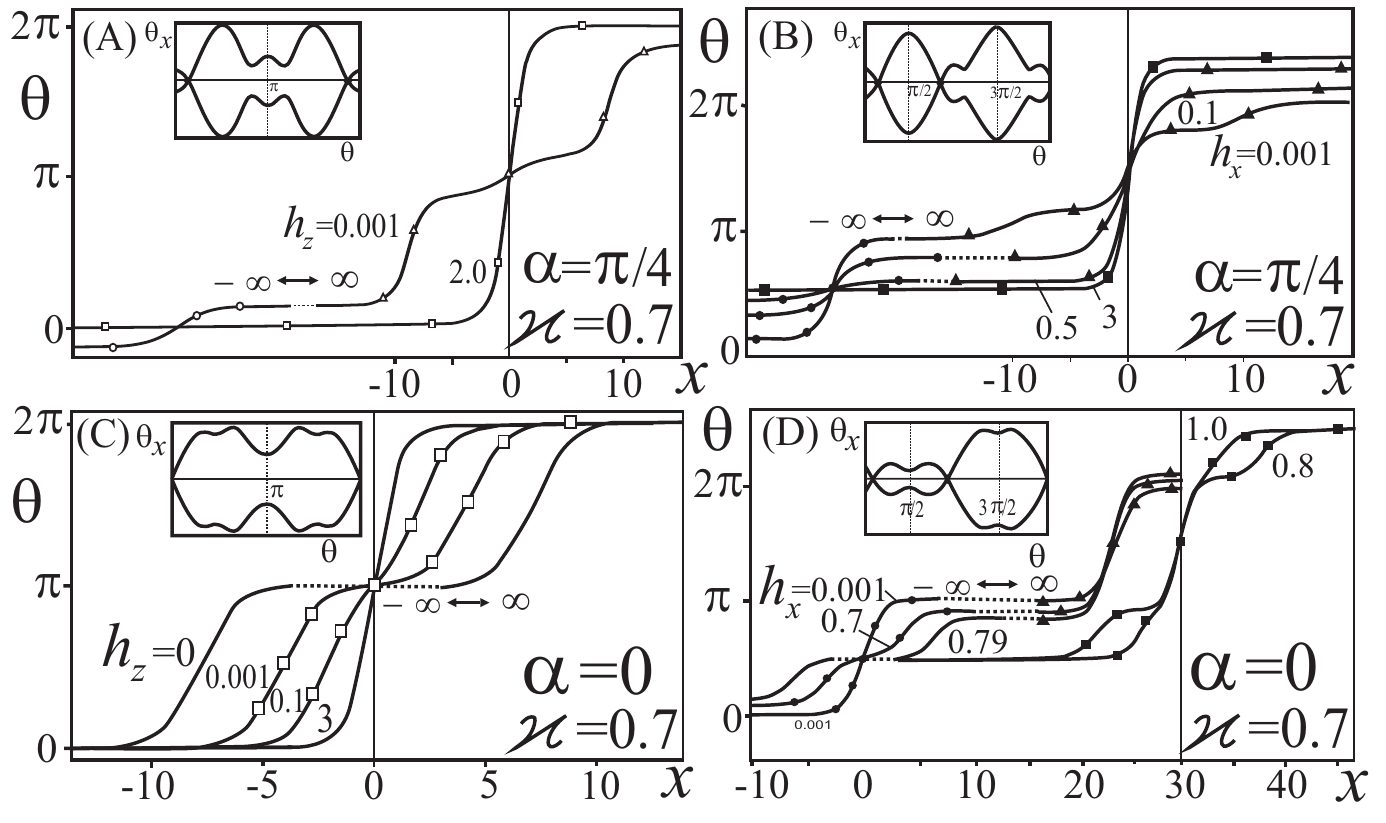}
\caption{
\label{walls1}
Domain wall profiles
for  $\varkappa =0.7$: 
(A), (B)- $\alpha =\pi/4$,
(C), (D) - $\alpha=0$.
Corresponding energy profiles are
plotted in Fig.~\ref{astroids2}.
Insets show the phase plane $(\theta;\theta_x)$
where $\theta_x=d\theta/dx$ (See Eq.(\ref{Euler})).%
}
\end{figure*}

In this case three types of domain walls can 
exists (Fig.~\ref{walls1}(A)):
DWI between domains with $\theta_1$, $\theta_2$ and
$\theta_3$, $\theta_4$
($\Delta \theta_{\textrm{I}}= |\theta_1-\theta_2| =
|\theta_3-\theta_4|= \arccos \varkappa$), 
DW II  between $\theta_1$, $\theta_3$ and
$\theta_2$, $\theta_4$
($\Delta \theta_{\textrm{II}}= |\theta_1-\theta_3| =
|\theta_2-\theta_4|= \pi - \Delta \theta_{\textrm{I}}$), and 
DWIII of $180^°$ type between domains $\theta_1$, $\theta_4$ and
$\theta_2$, $\theta_3$.
By integration of (\ref{Euler2}) 
the energy $\sigma$
and the magnetization profiles for the DWI
(upper signs) and DWII (lower signs) 
can be readily obtained as
\begin{equation}
\sigma=\delta_0\left[\sqrt{1-\varkappa^2}
\mp\varkappa\arccos(\pm \varkappa)\right]
\label{sigma1}
\end{equation}
\begin{equation}
x=\frac{x_0}{\sqrt{2(1-\varkappa^2)}}
\ln\left(\mp\frac{\tan\theta-\tan \theta_1}
{\tan\theta+\tan \theta_1}\right)\\,
\label{DWprofile}
\end{equation}
where $\delta_0=\sqrt{A\,K_c}$.
Depending on the ratio $1>\varkappa>0$
DWI with $\Delta \theta_I<90^°$
becomes more favourable than DWII 
with  $\Delta \theta_{II}>90^°$
and should exhibit stronger contrast 
in  experiment \cite{Sugawara07}.
Such domain boundaries 
 were experimentally
observed in thin films of (Ga,Mn)As by
Lorentz microscopy \cite{Sugawara07}.
From Ref.\onlinecite{Sugawara08}
and using (\ref{DWprofile}) we can evaluate
the DW width.
For DWI at the temperature $T=10\ (30)$K we obtain
$\delta=50\, (100)$nm while for DWII $\delta=43\, (76)$nm
which is consistent with the experiment.
Here, $A=0.4\cdot 10^{-12}\mathrm{Jm^{-1}}, 
K_c=1.18\,(0.32)\mathrm{Jm^{-3}}, 
K_u=0.18\,(0.11)\mathrm{Jm^{-3}}$.
For $\varkappa>1$ only DWIII exist.

These domain walls are characterized 
by strong variation of their parameters
with the applied magnetic field.
At magnetic field $h_z$ or $h_x$
domain walls of two types are present:  
walls where the magnetization vector rotates 
less or more than $180^0$
(Fig.\ref{walls1}(A),(B)).
Note, that for magnetic field $h_z>2(1-\varkappa)$
the metastable minima (Fig.\ref{astroids2}(B))
strongly influence the profile and energy
of the domain.
Such a remarkable modification of the structure
should strongly influence magnetoresistence
of domain walls (e. g. in nanoconstrictions).

In Fig.\ref{walls1}(C),(D) domain wall
profiles and typical phase portraits 
for the case $\alpha=0\ (\varkappa=0.7)$
are plotted.
At applied magnetic field $h_z$
only $360^°$ domain walls exist.
Within these walls, nuclei 
of the domain with $\pi$ and $\pm\pi/2$
are present
(see phase portrait in Fig.\ref{walls1}(C)).
In a magnetic field strong enough
these nuclei disappear.
So, the energy of domain wall increases,
although the width decreases.
By application of magnetic fields
perpendicular to the easy axis $\mathbf a$
in the interval $[0;a_1]$ there are walls 
of two types between upper and lower canted 
phases (Fig.\ref{walls1}(D)).
In each domain wall nuclei of domains with $\pi/2$ and
$3\pi/2$, correspondingly, are present.
At $h_x=h_x(a_1)$ these nuclei expand 
forming three-phase multidomain textures 
\cite{Titova05}.

\subsection{Parameters of stripe domains}

\begin{figure}[thb]
\includegraphics[width=8cm,keepaspectratio]{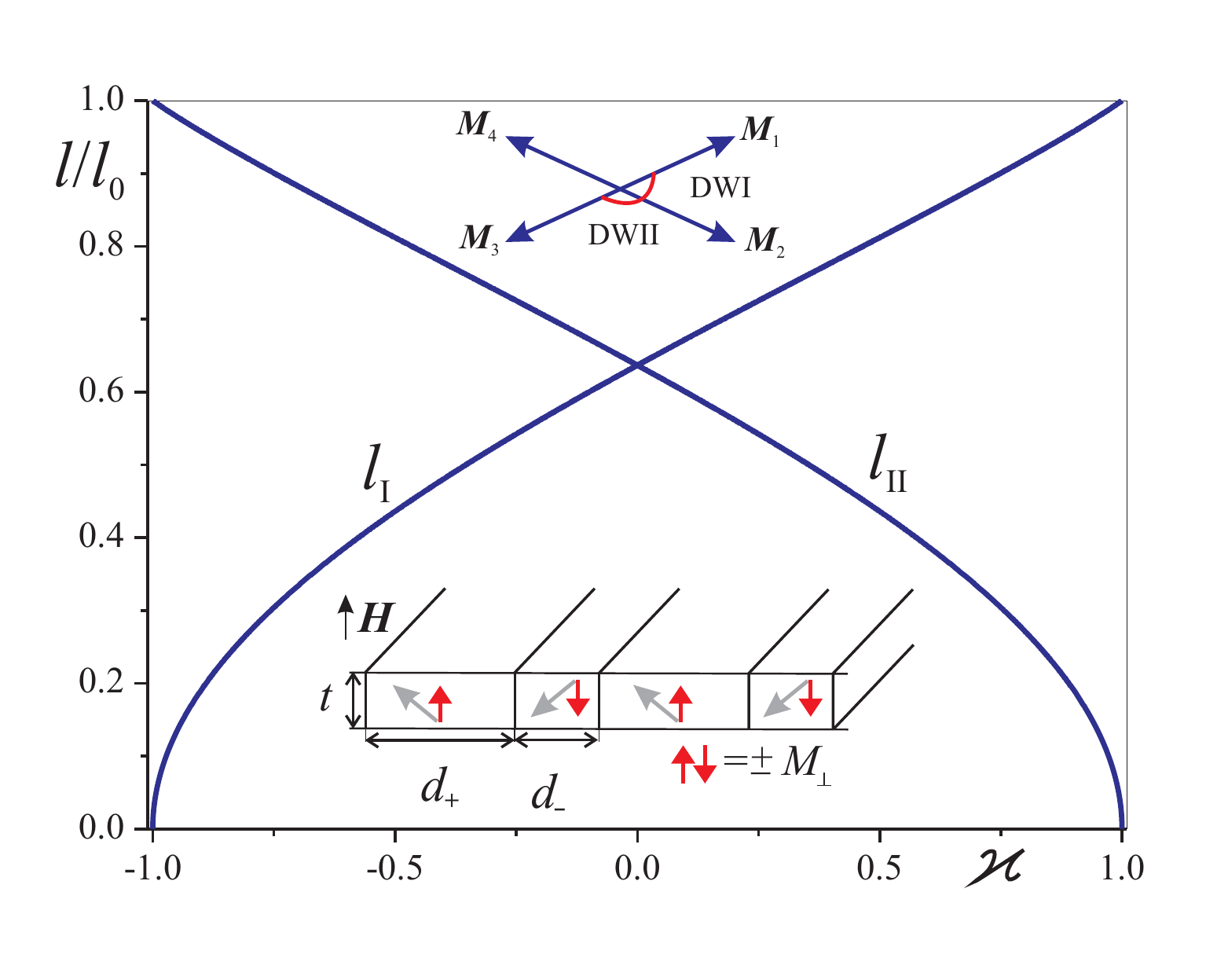}
\caption{
\label{stripes}
Characteristic lengths
$l_{I(II)}$
for stripe domains
with  DWI(DWII) in
four-phase state
($l_0 = \pi \sqrt{AK_c}/(4M_{0}^2$).
}
\end{figure}

Magnetic configurations in 
(Ga,Mn)As systems include a number of
noncollinear two- and multi-phase states.
These phases can create thermodynamically
stable multidomain states \cite{UFN88}.
For two coexisting
phases with the magnetization 
$\mathbf{M}^{(1)}$ and $\mathbf{M}^{(2)}$
effective values of magnetization
can be introduced \cite{UFN88}
\begin{align}
&M_{\perp} = (\mathbf{M}^{(1)}-\mathbf{M}^{(2)})\cdot \mathbf{v}/2,\nonumber\\
&H = \left[\mathbf{H} - \mathbf{H}_{tr}
- 4\pi (\mathbf{M}^{(1)}+\mathbf{M}^{(2)})
\right]\cdot \mathbf{v}/2.
\label{Mperp}
\end{align}
In particular, for  perpendicular magnetized 
(Ga,Mn)As nanolayers with $\varkappa > 5$
the phase diagrams of magnetic states 
(Fig. \ref{astroids1},(A),(E)) are similar to 
those for  uniaxial ferromagnets. 
In this case $M_{\perp}= M_0$ and domains
are separated  by 180$^\circ$ domain walls
\cite{Hubert98}.
According to Eq.( \ref{Mperp})
the problem of multidomain states for two-phase
noncollinear states can be reduced to a ferromagnetic 
collinear domain structure
with up and down magnetization $M_{\perp}$ in a bias
field $H$ ($\mathbf{v}$ is the unity vector perpendicular to
the layer plane,  $\mathbf{H}_{tr}$ is the transition
field between the phases 
$\mathbf{M}^{(1)}$and $\mathbf{M}^{(2)}$)
\cite{UFN88}.
Similar multidomain texture are formed in
the systems with a number of coexisting phases
larger than two.
For example, for $ -1 < \varkappa < 1$, $\alpha =0$
the magnetic configurations (Eq. (\ref{four}))
create four-phase multidomain states.
In a layer with $ \mathbf{a} || \mathbf{v}$
these textures  can be described by a model of
ferromagnetic domains with the magnetization
$M_{\perp} = M\sqrt{(1 +\kappa)/2}$.
(Fig. \ref{stripes}).

With effective values 
of the magnetization $M_{\perp}$
and bias field $H$  (\ref{Mperp})
the energy density
of a (Ga,Mn)As nanolayer with stripe
domains can be reduced to the well-studied
model of ferromagnetic stripes 
\cite{KooyEnz60,FTT80,Hubert98}
\begin{eqnarray}
w = 2 \pi M_{\perp}^2  \left[ w_m
+ \frac{2 \Lambda p}{\pi^2} - \frac{Hq}{2\pi M_{\perp}} \right]
\label{energyStripe}
\end{eqnarray}
with the stray field energy density given by
\begin{eqnarray}
w_m = 1- \frac{2p}{\pi^2} \int_0^1 (1-\tau)\ln 
\left[1 + \frac{ \cos^2 \left(\pi q/2 \right)}
{ \sinh^2 \left(p \tau/2 \right)}\right]d \tau\,,
\label{energyStripe2}
\end{eqnarray}
where $p = 2\pi t/D$, $ q = (d_{+}-d_{-})/D$,
$t$ is the layer of thickness $t$, 
$D = d_{+}+d_{-}$ is the stripe period,
and $d_{\pm}$ are domain sizes with up and down 
magnetization (Fig. \ref{stripes}).
The dimensionless parameter
$\Lambda =  \sigma(\varkappa)/(4M_{\perp}(\varkappa)^2t)= \pi l(\varkappa))/t$ 
measures the ratio between the domain wall energy $\sigma$
and the stray field energies.
It scales with the \textit{characteristic  length} 
$l(\varkappa) = \sigma/(4\pi M_{\perp}^2)$
as the relevant material parameter \cite{Hubert98}.
Minimization of (\ref{energyStripe}) 
with respect to $p$ and $q$
derives the solutions for the geometric 
parameters  $d_{\pm}$ as functions
of three control parameters of
model (\ref{energyStripe2}):
the layer thickness $t$, 
the bias field $H$,
and factor $\Lambda$.
These solution have been investigated
in detail (see  \cite{KooyEnz60,FTT80,
Hubert98} and bibliography
in Ref. \cite{Hubert98}).
Particularly, it was shown that
the solutions for stripes 
exist only below certain critical 
field $H < H^*(\Lambda) < 4 \pi M_{\perp}$
\cite{KooyEnz60}. 
As the bias field approaches $H^*$ 
the stripes gradually
transform into the homogeneous
state by unlimited growth of the
period ($D \rightarrow \infty$).
However, at the critical
field the domain of the minority
phase preserves a finite size
$d_{-}(H^*) = d_{-}^*$. 
At higher fields 
($H^* > H > 4\pi M_{\perp}$)
it exists
as a metastable state
gradually shrinking  to zero size
at the saturation field.

In perpendicular
magnetized (Ga,Mn)As layers
the period of multidomain patterns
exceed their thicknesses 
\cite{Shono00,Thevenard06,Gourdon07}.
For such large stripes ($D\geq t$)
the expansion of the integral
(\ref{energyStripe2}) allows
to simplify the problem
 \cite{FTT80}. 
After some algebra, the
solutions for stripes can be
derived in analytical form
as a set of parametrical
equations 
\begin{align}
&D(H) =\pi u t /\sqrt{1 -\left(H/H^* \right)^2 },\nonumber\\
&d_{\pm}= (D/\pi)\arccos (\mp H/H^*),
\label{solStripes}
\end{align}%
\begin{eqnarray}
H^* (u)=4\pi M_{\perp} f(u), 
\quad
2 \Lambda = g(u).
\label{trans1}
\end{eqnarray}%
Here we introduce parameter
$u = d_{-}^*/t$ and  functions
\begin{eqnarray}
f(u) = \left[2\arctan{1/u}-u\ln(1+1/u^2)\right]/ \pi, 
 \nonumber\\
g(u) = (1+u^2)\ln(1+u^2)-u^2\ln(u^2). 
\label{trans2}
\end{eqnarray}%
According to (\ref{solStripes}) at zero
field $D(0)= D_0 = \pi t u$, thus,
 the ratio $D_0/d_{-}^* = \pi$.
It means that at the transition field
the domain of the minority phase
becomes approximately six times 
narrower than the domain size
at zero field ($D_0/2$).
Within this approximation 
the equilibrium magnetization in
the stripe phase equals

\begin{eqnarray}
\left\langle M \right\rangle = M_{\perp} q 
= (2 M_{\perp}/\pi) \arccos (H/H^*).
\label{solStripesb}
\end{eqnarray}%

Finally, in the limit of large domains
$D \gg t$,  $D_0 = \pi t \exp (\Lambda -1/2)$,
and the transition fields for
stripe and bubble domains 
becomes exponentially small,
e. g. transition field
$H^* = 4 M_{\perp} \exp (- \Lambda +1/2)$,
 the bubble collapse field
$H_c = 16 M_{\perp} \exp (- \Lambda -1/2)$,
and ratio $H^*/H_c = e/4 =0.6796 $
\cite{FTT80}.

These results
show that the solutions
of magnetic domains
in (Ga,Mn)As layers should demonstrate
general features similar 
to those in uniaxial ferromagnets.
However, there is an important difference
between these two systems.
In uniaxial ferromagnets the characteristic
length is expressed as a combination of basic
magnetic parameters (constants of uniaxial anisotropy
$K$, exchange stiffness $A$ and saturation
magnetization $M$):
$l_f = \sigma_f/(4 \pi M^2) = \sqrt{AK}/(\pi M^2)$
depends only on values of uniaxial anisotropy $K$.
On the contrary, in the diluted magnetic semiconductors
the characterstic length $l(\varkappa)$ strongly
depends on the values of competing magnetic anisotropies
and varies in a broad range providing a complex behaviour
of multidomain patterns in these materials.
Eqs. (\ref{solStripes}), (\ref{trans1}),
(\ref{solStripesb}) connect 
equilibrium parameters of
stripe domains with material
parameters of (Ga,Mn)As
systems (\ref{potential}).
For this model 
calculations of $M_{\perp}$
and the domain wall energy
$\sigma$ allow to express
$\Lambda$ as a function
of $\kappa$ and $\alpha$.
For example, for four-phase
domains (Eq. (\ref{four}))
the magnetization
and the domain wall energies 
(Eq. (\ref{sigma1})) yield
functions $l_{I(II)} (\varkappa)$
plotted in Fig. \ref{stripes}.

Stripe domains have been observed in a number
of (Ga,Mn)As nanolayers
\cite{Shono00,Fukumura01,
Pross04,Thevenard06,Shin07,
Thevenard07,Dourlat07,Gourdon07}.
In Table I we collect experimental
data (indicated by bold) and the calculated
 stripe domains parameters
(by solving Eqs. (\ref{solStripes}), (\ref{trans1}),
(\ref{solStripesb})) for (Ga,Mn)As layers
(a-c), an yttrium-iron garnet film (d) and
FePd nanolayers (e,f).
For thin layers of $\mathrm{(Ga_{0.957}Mn_{0.043})As}$ 
for $T$ = 9 K (1) and $T$ = 80 K (2)
\cite{Shono00} we use the experimental values
of $t$ and $D_0$ to calculate other parameters
of stripes.
For a layer $\mathrm{(Ga_{0.93}Mn_{0.07})As}$
at $T$ = 80 K \cite{Gourdon07} we use $t$ 
and a value of the transition field $H^*/(4 \pi M_{\perp})$
to calculate $l$, $\Lambda$ and domain sizes
$D_0$ and $d_{-}^*$.
For comparison  we derive stripe domain parameters for
an epitaxial garnet film 
$\mathrm{Y_{1.88}Lu_{0.2}Ca_{0.92}Ge_{0.92}Fe_{4.08}O_{12}}$
\cite{Bobeck}, and for FePd nanolayers \cite{Gehanno97}.
According to \onlinecite{Gourdon07} the saturation magnetization
in $\mathrm{(Ga_{0.93}Mn_{0.07})As}$   $M_s$ = 28 kA/m.
Then, from the calculated value of 
the characteristic length we  derive  
$\sigma$ = 49.0 $\mu$J/m$^2$ (for comparison, 
in the (Y,Fe) garnet film $M_s$ = 13.6 kA/m,
and $\sigma$ = 110  $\mu$J/m$^2$ \cite{Bobeck}).
In garnet films and other classical materials with 
perpendicular anisotropy  regular stripe domains 
are observed, if the layer thickness is considerably larger
than the characterisitic length. 
In such systems 
the equilibrium domain sizes at zero field
$D_0$ do not exceed the layer thicknesses
($D_0 \leq t$).
When films becames
thinner than $l$ (e.g. in the vicinity of
the compensation temperature of ferrimagnets
\cite{FTT80}) the demagnetization
forces are too weak to overcome coercivity
and the formation  equilibrium domains
is impeded.
As a result such multidomain patterns consist of
very large domains with irregular 
boundaries \cite{Hubert98}.
Similar disordered domains 
and strongly hysteretic
behaviour have been observed
in (Ga,Mn)As films
with perpendicular anisotropy
\cite{Shono00,Pross04,
Thevenard07,Dourlat07,Gourdon07}.
It means that the equilibium multidomain
state are hardly reached in these
systems.
For example, in $\mathrm{(Ga_{0.93}Mn_{0.07})As}$
layers the observed width of the minority
stripe at the critical field is
$d_{-}^* = 1.7 \pm 0. 3$ $\mu$m \cite{Gourdon07},
and about one order larger than the calculated
equilibrium value ($d_{-}^* = 0.2 $ $\mu$m).
%
%
Up to now only few results
on experimental investigations of
multidomain states in perpendicularly
magnetized (Ga,Mn)As nanolayers
have been reported.
New detailed investigations involving
modern experimental methods developed
in other fields of nanomagnetism 
(see e.g. \onlinecite{Hellwig07}) 
are desirable.
The results of this section

establish important physical connections
with multidomain states in other classes
of perpendicular magnetized materials
and provide a theoretical basis
for future research.

\begin{table*}[thb]
\caption{Parameters of the stripe domains in (Ga,Mn)As layers
(a-c), a (Y,Fe) garnet film (d) and FePd nanolayers (e,f). 
They include experimental values (bold) and the results
derived from model (\ref{energyStripe}).
Here, $t$ is the layer thickness, $l$ is the characteristic 
length, $\Lambda = \pi l/t$ is the dimensionless
parameter measuring the ratio between the domain wall
energy and the stray field energy (see Eqs. (22),(23))),
$D_0$ is the equilibrium period at zero field, 
$d_{-}^{*}$ is the equilibrium size of the minority phase
at $H= H^{*}$, the trasition field into the homogeneous 
state, Eq. (25).
 }
\label{tab1}
\centering
\begin{tabular}{ccccccc}
 & $t$, $\mu$m & \hspace{6mm}
\hspace{1mm} $l$, $\mu$m & \hspace{2mm} $\Lambda$ \hspace{4mm} &   \hspace{6mm}$D_0$, $\mu$m \hspace{2mm} &   \hspace{4mm} $d^{*}_{-}$,$\mu$m \hspace{2mm} &   \hspace{3mm} $H^{*}/(4 \pi M_{\perp})$  \\
\hline
(a). $\mathrm{(Ga_{0.957}Mn_{0.043})As}$ \cite{Shono00}& \textbf{0.2} & 0.132  &  2.07 & \textbf{3.0} & 0.95 & 6.6 10$^{-2}$ \\
\hline
(b). $\mathrm{(Ga_{0.957}Mn_{0.043})As}$ \cite{Shono00} & \textbf{0.2} & 0.220  & 3.45 & \textbf{12.0} & 3.82 & 1.7 10$^{-2}$  \\
\hline
(c).  $\mathrm{(Ga_{0.93}Mn_{0.07})As}$ \cite{Gourdon07} & \textbf{5 10$^{-2}$} & 0.10 & 1.920  & 0.643 & 0.2038 & \textbf{3.9 10$^{-2}$}\\
\hline
(d).  (Y,Fe) garnet film \cite{Bobeck} & \textbf{11.0} & \textbf{0.47} & 0.1342  & 12.50 & 3.0 & 0.6  \\
\hline
(e).  FePd nanolayer \cite{Gehanno97}  & \textbf{3.6 10$^{-2}$}& 9 10$^{-3}$ & 0.7526 & \textbf{0.13} & 3.9 10$^{-2}$ & 0.2612 \\
\hline 
(f).   FePd nanolayer \cite{Gehanno97} & \textbf{1.15 10$^{-2}$} & 9 10$^{-3}$ & 2.477 & \textbf{0.26} & 8.3 10$^{-2}$ & 4.4 10$^{-2}$  \\
\end{tabular}
\end{table*}

\section{Summary and Conclusions}

We have developed some micromagnetic
methods \cite{UFN88,Hubert98} which give
a consistent description of magnetization
processes and multidomain structure
in systems with competing
anisotropies such as diluted magnetic
semiconductors.

Theoretically constructed phase diagrams
in external field components in the limiting 
case of ideally soft magnetics allow to
understand the creation of equilibrium 
domain structure \cite{Welp03},
within ahysteretic magnetization reversal,
and explain various parts of magnetization
curves (Fig.\ref{curves}).
Thus, magnetic phase diagrams 
allow to put in good order and classify
a vast amount of experimental data
on reorientation effects,
multidomain processes, and
magnetization reversal 
in (Ga, Mn) As systems. 
These diagrams also give opportunity to
predict change of magnetic states of the
system in zero magnetic field under action
of temperature \cite{Shono00,Fukumura01,Pross04}
and in the 
case of arbitrary angle $\alpha$ between
competing anisotropy axes \cite{311A,311B}.

It is also shown that
the applied magnetic field
causes drastic transformations of
the domain wall profile 
and strongly influences its parameters.
Domain walls can serve as nuclei of
domains for a new phase.
At certain values of the magnetic field
a domain wall can be divided into 
domains of a new phase and two types
of new domain walls.
At certain critical endpoints of
phase coexistence, domain walls
can disappear by the rotation of
the magnetization in adjacent
domains towards each other.

For nanolayers with perpendicular anisotropy 
the geometrical parameters of stripe domains 
have been calculated as functions of a bias field.

\begin{acknowledgments}

The authors thank
J. McCord, V. Neu, 
and R. Sch{\"a}fer for helpful discussions.
%
%
Work supported by DFG through SPP1239, project A8.
A.A.L. and A.N.B.\ thank H.\ Eschrig for support and
hospitality at IFW Dresden. 
\end{acknowledgments}

\end{document}